\documentclass[double-spaced]{llncs}

\usepackage{amssymb}
\usepackage{xspace}
\usepackage[english]{babel}
\usepackage{amsmath}
\usepackage{amsopn}
\usepackage{latexsym}
\usepackage{textcomp}
\usepackage{epsfig}
\usepackage{subfigure}
\usepackage{enumerate}
\usepackage{tabularx}
\usepackage[plainpages=false]{hyperref}
\usepackage[figure,table]{hypcap}
\usepackage{multirow}
\usepackage{todonotes}
\usepackage{color, colortbl}

\usepackage{amsmath}
\usepackage{algorithm}
\usepackage[noend]{algpseudocode}
\makeatletter
\def\BState{\State\hskip-\ALG@thistlm}
\makeatother

\newcommand{\remove}[1]{}
\newcommand{\spinner}{\textsc{Spinner}\xspace}
\newcommand{\fmmm}{\textsc{FM3}\xspace}
\newcommand{\multi}{\textsc{Multi-GiLA}\xspace}
\newcommand{\gila}{\textsc{GiLA}\xspace}
\newcommand{\regularb}{\textsc{RegularGraphs}\xspace}
\newcommand{\realb}{\textsc{RealGraphs}\xspace}
\newcommand{\bigb}{\textsc{BigGraphs}\xspace}

\title{A Distributed Multilevel Force-directed Algorithm%
\thanks{Research supported in part by the MIUR project AMANDA ``Algorithmics for MAssive and Networked DAta'', prot. 2012C4E3KT\_001.}
}
\author{
Alessio Arleo\inst{1},
Walter Didimo\inst{1},
Giuseppe Liotta\inst{1},
Fabrizio Montecchiani\inst{1},
}

\date{}

\institute{
Universit\`a degli Studi di Perugia, Italy\\
\email{alessio.arleo@studenti.unipg.it} \\
\email{\{walter.didimo,giuseppe.liotta,fabrizio.montecchiani\}@unipg.it}
}

\begin{document}
\maketitle

\begin{abstract}
The wide availability of powerful and inexpensive cloud computing services naturally motivates the study of distributed graph layout algorithms, able to scale to very large graphs. Nowadays, to process Big Data, companies are increasingly relying on PaaS infrastructures rather than buying and maintaining complex and expensive hardware. So far, only a few examples of \emph{basic} force-directed algorithms that work in a distributed environment have been described. Instead, the design of a distributed \emph{multilevel} force-directed algorithm is a much more challenging task, not yet addressed. We present the first multilevel force-directed algorithm based on a distributed vertex-centric paradigm, and its implementation on Giraph, a popular platform for distributed graph algorithms. Experiments show the effectiveness and the scalability of the approach. Using an inexpensive cloud computing service of Amazon, we draw graphs with ten million edges in about $60$ minutes.   
\end{abstract}

\section{Introduction}\label{se:introduction}

Force-directed algorithms are very popular techniques to automatically compute graph layouts. They model the graph as a physical system, where attractive and repulsive forces act on each vertex. Computing a drawing corresponds to finding an equilibrium state (i.e., a state of minimum energy) of the force system through a simple iterative approach. Different kinds of force and energy models give rise to different graph drawing algorithms. Refer to the work of Kobourov for a survey on the many force-directed algorithms described in the literature~\cite{Handbook-kob}. Although basic force-directed algorithms usually compute nice drawings of small or medium graphs, using them to draw large graphs has two main obstacles: $(i)$ There could be several local minima in their physical models: if the algorithm falls in one of them, it may produce bad drawings. The probability of this event and its negative effect increase with the size of the graph. $(ii)$ Their approach is computationally expensive, thus it gives rise to scalability problems even for graphs with a few thousands of vertices.

To overcome the above obstacles, \emph{multilevel} force-directed algorithms have been conceived. A limited list of works on this subject includes~\cite{DBLP:journals/isci/DidimoM14}, \cite{DBLP:journals/comgeo/GajerGK04}, \cite{DBLP:conf/gd/HachulJ04}, \cite{DBLP:journals/dam/HadanyH01,DBLP:journals/jgaa/HarelK02}, \cite{Hu-05}, \cite{DBLP:journals/jgaa/Walshaw03} (see~\cite{Handbook-kob} for more references). These algorithms generate from the input graph $G$ a series (hierarchy) of progressively simpler structures, called \emph{coarse graphs}, and then incrementally compute a drawing of each of them in reverse order, from the simplest to the most complex (corresponding to $G$).  On common machines, multilevel force-directed algorithms perform quickly on graphs with several thousand vertices and usually produce qualitatively better drawings than basic algorithms~\cite{DBLP:conf/gd/BartelGKM10}, \cite{JGAA-150}, \cite{Handbook-kob}. Implementations based on GPUs have been also experimented~\cite{DBLP:conf/gd/GodiyalHGH08}, \cite{DBLP:journals/tvcg/IngramMO09}, \cite{skssl-sunc+-2011}, \cite{yya-sffg+-12}. They scale to graphs with a few million edges, but their development requires a low-level implementation and the necessary infrastructure could be expensive in terms of hardware and maintenance.

The wide availability of powerful and inexpensive cloud computing services and the growing interest towards PaaS infrastructures observed in the last few years, naturally motivate the study of distributed graph layout algorithms, able to scale to very large graphs. So far, the design of distributed graph visualization algorithms has been only partially addressed. Mueller \emph{et al.}~\cite{DBLP:conf/egpgv/MuellerGL06} and Chae \emph{et al.}~\cite{csmag-hdgv+-2012} proposed force-directed algorithms that use multiple large displays. Vertices are evenly distributed on the different displays, each associated with a different processor, which is responsible for computing the positions of its vertices; scalability experiments are limited to graphs with some thousand vertices. Tikhonova and Ma~\cite{DBLP:conf/egpgv/TikhonovaM08} presented a parallel force-directed algorithm that can run on graphs with few hundred thousand edges. It takes about $40$ minutes for a graph of $260,385$ edges, on $32$ processors of the PSC's BigBen Cray XT3 cluster. More recently, the use of emerging frameworks for distributed graph algorithms has been investigated. Hinge and Auber~\cite{DBLP:conf/iv/AntoineD15} described a distributed basic force-directed algorithm implemented in the Spark framework, using the GraphX library. Their algorithm is mostly based on a MapReduce paradigm and shows margins for improvement: it takes $5$ hours on a graph with $8,000$ vertices and $35,000$ edges, on a cluster of $16$ machines, each equipped with $24$ cores and $48$ GB of RAM. A distributed basic force-directed algorithm running on the Apache Giraph framework has been presented in~\cite{DBLP:conf/gd/ArleoDLM15} (see also~\cite{arxiv} for an extended version of this work). Giraph is a popular platform for distributed graph algorithms, based on a vertex-centric paradigm, also called the TLAV (``Think Like a Vertex'') paradigm~\cite{c-glgp+-2011}. Giraph is used by Facebook to analyze the huge network of its users and their connections~\cite{DBLP:journals/pvldb/ChingEKLM15}. The algorithm in~\cite{DBLP:conf/gd/ArleoDLM15} can draw graphs with a million edges in a few minutes, running on an inexpensive cloud computing infrastructure. However, the design of a distributed multilevel force-directed algorithm is a much more challenging task, due to the difficulty of efficiently computing the hierarchy required by a multilevel approach in a distributed manner (see, also~\cite{arxiv}, \cite{DBLP:conf/iv/AntoineD15}).

\paragraph{Our Contribution.} This paper presents \multi (Multilevel Giraph Layout Algorithm), the first distributed multilevel force-directed algorithm based on the TLAV paradigm and running on Giraph. The model for generating the coarse graph hierarchy is inspired by \fmmm (Fast Multipole Multilevel Method), one of the most effective multilevel techniques described in the literature~\cite{DBLP:conf/gd/BartelGKM10}, \cite{DBLP:conf/gd/HachulJ04,JGAA-150}. 
The basic force-directed algorithm used by \multi to refine the drawing of each coarse graph is the distributed algorithm in~\cite{arxiv} (Section~\ref{se:multigila}). We show the effectiveness and the efficiency of our approach by means of an extensive experimental analysis: \multi can draw graphs with ten million edges in about $60$ minutes (see Section~\ref{se:experiments}), using an inexpensive PaaS of Amazon, and exhibits high scalability. To allow replicability of the experiments, our source code and graph benchmarks are made publicly available~\cite{multi}.
It is worth observing that in order to get an overview of the structure of a very large graph and subsequently explore it in more details, one can combine the use of \multi with systems like \textsc{LaGo}~\cite{DBLP:journals/tvcg/ZinsmaierBDS12}, which provides an interactive level-of-detail rendering, conceived for the exploration of large graphs (see Section~\ref{se:experiments}). Section~\ref{se:preliminaries} contains the necessary background on multilevel algorithms and on Giraph. Conclusions and future research are in Section~\ref{se:future}. Additional figures can be found in the appendix.

\section{Background}\label{se:preliminaries}

{\noindent \bf Multilevel force-directed algorithms.} Multilevel force-directed algorithms work in three main phases: \emph{coarsening}, \emph{placement}, and \emph{single-level layout}. Given an input graph $G$, the coarsening phase computes a sequence of graphs $\{G=G_0, G_1,\dots, G_k\}$, such that the size of $G_{i+1}$ is smaller than the size of $G_i$, for $i=0,\dots,k-1$. To compute $G_{i+1}$, subsets of vertices of $G_i$ are merged into single vertices. The criterion for deciding which vertices should be merged is chosen as a trade-off between two conflicting goals. On one hand, the overall graph structure should be preserved throughout the sequence of graphs, as it influences the way the graph is unfolded. On the other hand, both the number of graphs in the sequence and the size of the coarsest graph may have a significant influence on the overall running time of the algorithm.  Therefore, it is fundamental to design a coarsening phase that produces a sequence of graphs whose sizes quickly decrease, and, at the same time, whose structures smoothly change. The sequence of graphs produced by the coarsening phase is then traversed from $G_k$ to $G_0=G$, and a final layout of $G$ is obtained by progressively computing a layout for each graph in the sequence.  In the placement phase, the vertices of $G_i$ are placed by exploiting the information of the (already computed) drawing $\Gamma_{i+1}$ of $G_{i+1}$. Starting from this initial placement, in the single-level (basic) layout phase, a drawing $\Gamma_i$ of $G_i$ is computed by applying a single-level force-directed algorithm. Thanks to the good initial placement, such an algorithm  will reach an equilibrium after a limited number of iterations. For $G_k$ an initial placement is not possible, thus the layout phase is directly applied starting from a random placement.

Since our distributed multilevel force-directed algorithm is partially based on the \fmmm algorithm, we briefly recall how the coarsening and placement phases are implemented by \fmmm (see~\cite{DBLP:phd/de/Hachul2005,DBLP:conf/gd/HachulJ04} for details). Let $G=G_0$ be a connected graph (distinct connected components can be processed independently), the coarsening phase is implemented through the \textsc{Solar Merger} algorithm. The vertices of $G$ are partitioned into vertex-disjoint subgraphs called \emph{solar systems}. The diameter of each solar system is at most four. Within each solar system $S$, there is a vertex $s$ classified as a \emph{sun}. Each vertex $v$ of $S$ at distance one (resp., two) from $s$ is classified as a \emph{planet} (resp., a \emph{moon}) of $S$. There is an \emph{inter-system link} between two solar systems $S_1$ and $S_2$, if there is at least an edge of $G$ between a vertex of $S_1$ and a vertex of $S_2$. The coarser graph $G_1$ is obtained by collapsing each solar system into the corresponding sun, and the inter-system links are transformed into edges connecting the corresponding pairs of suns. Also, all vertices of $G=G_0$ are associated with a \emph{mass} equal to one. The mass of a sun is the sum of the masses of all vertices in its solar system.  The coarsening procedure halts when a coarse graph has a number of vertices below a predefined threshold. The placement phase of \fmmm is called \textsc{Solar Placer} and uses information from the coarsening phase. The vertices of $G_{i+1}$ correspond to the suns of $G_i$, whose initial position is defined in the drawing $\Gamma_{i+1}$. The position of each vertex $v$ in $G_{i} \setminus G_{i+1}$ is computed by taking into account all inter-system links to which $v$ belongs. The rough idea is to position $v$ in a barycentric position with respect to the positions of all suns connected by an inter-system link that passes through $v$. 

{\noindent \bf The TLAV paradigm and the Giraph framework.} The TLAV paradigm requires to implement distributed algorithms from the perspective of a vertex rather than of the whole graph. Each vertex can store a limited amount of data and can exchange messages only with its neighbors. The TLAV framework Giraph~\cite{c-glgp+-2011} is built on the Apache Hadoop infrastructure and originated as the open source counterpart of Google's \emph{Pregel}~\cite{mabd+-pslgp-2010} (based on the \emph{BSP model}~\cite{v-bmpc-1990}). In Giraph, the computation is split into \emph{supersteps} executed iteratively and synchronously. A superstep consists of two phases: $(i)$ Each vertex executes a user-defined vertex function based on both local vertex data and on data coming from its adjacent vertices; $(ii)$ Each vertex sends the results of its local computation to its neighbors, along its incident edges. The whole computation ends after a fixed number of supersteps or when certain user-defined conditions are met (e.g., no message has been sent or an equilibrium state is reached).

{\noindent \bf Design challenges and the \gila algorithm.}  Force-directed algorithms (both single-level and multilevel) are conceived as sequential, shared-memory graph algorithms, and thus are inherently centralized. On the other hand, the following three properties must be guaranteed in the design of a TLAV-based algorithm: {\tt P1.} Each vertex exchange messages only with its neighbors; {\tt P2.} Each vertex locally stores a small amount of data; {\tt P3.} The communication load in each supertsep (number and length of messages sent in the superstep) is small: for example, linear in the number of edges of the graph. Property {\tt P1} corresponds to an architectural constraint of Giraph. Violating {\tt P2} causes out-of-memory errors during the computation of large instances, which translates in the impossibility of storing large routing tables in each vertex to cope with the absence of global information. Violating {\tt P3} quickly leads to inefficient computations, especially on graphs that are locally dense or that have high-degree vertices. Hence, sending heavy messages containing the information related to a large part of the graph is not an option.   

In the design of a multilevel force-directed algorithm, the above three constraints {\tt P1}--{\tt P3} do not allow for simple strategies to make a vertex aware of the topology of a large part of the graph, which is required in the coarsening phase. In Section~\ref{se:multigila} we describe a sophisticated distributed protocol used to cope with this issue. For the same reason, a vertex is not aware of the positions of all other vertices in the graph, which is required to compute the repulsive forces acting on the vertex in the single-level layout phase. The algorithm described in~\cite{arxiv}, called \gila, addresses this last issue by adopting a locality principle, based on the experimental evidence that in a drawing computed by a force-directed algorithm (see, e.g.,~\cite{Handbook-kob}) the graph theoretic distance between two vertices is a good approximation of their geometric distance, and that the repulsive forces between two vertices $u$ and $v$ tend to be less influential as their geometric distance increases. Following these observations, in the \gila algorithm, the resulting force acting on each vertex $v$ only depends on its \emph{$k$-neighborhood} $N_v(k)$, i.e., the set of vertices whose graph theoretic distance from $v$ is at most $k$, for a predefined small constant $k$. Vertex $v$ acquires the positions of all vertices in $N_v(k)$ by means of a controlled flooding technique. 
According to an experimental analysis in~\cite{arxiv}, $k=3$ is a good trade-off between drawing quality and running time. The attractive and repulsive forces acting on a vertex are defined using Fruchterman-Reingold model~\cite{fr-gdfdp-91}.

\section{The \multi Algorithm}\label{se:multigila}

In this section we describe our multilevel algorithm \multi. It is designed having in mind the challenges and constraints discussed in Section~\ref{se:preliminaries}. The key ingredients of \multi are a distributed version of both the \textsc{Solar Merger} and of the \textsc{Solar Placer} used by \fmmm, together with a suitable dynamic tuning of \gila. 

\subsection{Algorithm overview}

The algorithm is based on the pipeline described below. The pruning, partitioning, and reinsertion phases are the same as for the \gila algorithm, and hence they are only briefly recalled (see~\cite{arxiv} for details).

\smallskip\noindent{\bf Pruning:}  In order to lighten the algorithm execution, all vertices of degree one are temporarily removed from the graph; they will be reinserted at the end of the computation by means of an ad-hoc technique.

\smallskip\noindent{\bf Partitioning:} The vertex set is then partitioned into subsets, each assigned to a computing unit, also called \emph{worker} in Giraph (each computer may have more than one worker). The default partitioning algorithm provided by Giraph may create partitions with a very high number of edges that connect vertices of different partition sets; this would negatively affect the communication load between different computing units. To cope with this problem, we use a partitioning algorithm by Vaquero {\em et al.}~\cite{DBLP:conf/icdcs/VaqueroCLM14}, called \spinner, which creates balanced partition sets by exploiting the graph topology. 

\smallskip\noindent{\bf Layout:} This phase executes the pipeline of the multilevel approach. The coarsening phase (Section~\ref{sse:coarsening}) is implemented by means of a distributed protocol, which attempts to behave as the \textsc{Solar Merger} of \fmmm. The placement (Section~\ref{sse:placing}) and single-level layout (Section~\ref{sse:layout}) phases are iterated until a drawing of the graph is computed. 

\smallskip\noindent{\bf Reinsertion:} For each vertex $v$, its neighbors of degree one (if any) are suitably reinserted in a region close to $v$, avoiding to introduce additional edge crossings. 

This pipeline is applied independently to each connected component
of the graph, and the resulting layouts are then arranged in a matrix to avoid overlaps.

\subsection{Coarsening Phase: \textsc{Distributed Solar Merger}}\label{sse:coarsening}

Our \textsc{Distributed Solar Merger} algorithm yields results (in terms of number of levels) comparable to those obtained with the \textsc{Solar Merger} of \fmmm (see also Section~\ref{se:experiments}). The algorithm works into four steps described below; each of them involve several Giraph supersteps. For every iteration $i$ of these four steps, a new coarser graph $G_i$ is generated, until its number of vertices is below a predefined threshold. We use the same terminology as in Section~\ref{se:preliminaries}, and equip each vertex with four properties called \emph{ID}, \emph{level}, \emph{mass}, and \emph{state}. The ID is the unique identifier of the vertex. The level represents the iteration in which the vertex has been generated. That is, a vertex has level $i$ if it belongs to graph $G_i$. The vertices of the input graph have level zero. The second property represents the mass of the vertex and it is initialized to one  plus the number of its previously pruned neighbors of degree one for the vertices of the input graph. The state of a vertex can receive one of the following values: \emph{sun}, \emph{planet}, \emph{moon}, or \emph{unassigned}. We shall call sun, planet, moon, or unassigned, a vertex with the corresponding value for its state. All vertices of the input graph are initially unassigned. 

\begin{figure}[t]
\centering
\subfigure[]{\includegraphics[scale=0.9,page=1]{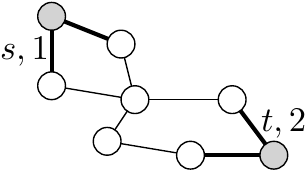}\label{fi:co-1}}\hfil
\subfigure[]{\includegraphics[scale=0.9,page=2]{figures/example}\label{fi:co-2}}\hfil
\subfigure[]{\includegraphics[scale=0.9,page=3]{figures/example}\label{fi:co-3}}
\caption{\label{fi:co}Illustration for the coarsening phase. (a) Two suns $s$ (ID 1) and $t$ (ID 2) broadcast an offer message. (b) The dark gray vertices receive the offer messages, become planets, and forward the received offer messages. The striped vertex will then receive offer messages from both $s$ and $t$, and (c) will accept the offer message of $t$ due to the greatest ID of $t$. In (c) the final galaxies are enclosed by dashed curves, suns (planets, moons) are light gray (dark gray, black).}
\end{figure}

\smallskip\noindent{\bf Sun Generation.} In the first superstep, each  vertex turns its state to sun with probability $p$, for a predefined value of $p$. The next three supersteps aim at avoiding pairs of suns with graph theoretic distance less than 3. First, each sun broadcasts a message containing its ID. In the next superstep, if a sun $t$ receives a message from an adjacent sun $s$, then also $s$ receives a message from $t$, and the sun between $s$ and $t$ with lower ID changes its state to unassigned. In the same superstep, all vertices (of any state) broadcast to their neighbors only the messages received from those vertices still having state sun. In the third superstep, if a sun $t$ receives a message generated from a sun $s$ (with graph theoretic distance 2 from $t$), again also $t$ receives a message from $s$ and the sun with lower ID changes its state to unassigned. This procedure ensures that all pairs of suns have graph theoretic distance at least three.  

\smallskip\noindent{\bf Solar System Generation.} In the first superstep, each sun broadcasts an \emph{offer message}. At the next superstep, if an unassigned vertex $v$ receives an offer message $m$ from a sun $s$, then $v$ turns its state to planet and stores the ID of $s$ in a property called  \emph{system-sun}.  Also, $v$ sends a \emph{confirmation message} to $s$. Finally, $v$ forwards the message $m$ to all its neighbors. At the next superstep, every sun vertex processes the received confirmation messages. If a sun $s$ received a confirmation message, $s$ stores the ID of the sender in a property called \emph{planet-list}. This property is used by each sun to keep track of the planets in its solar system. If a planet $v$ receives an offer message, then such a message comes from the same sun stored in the system-sun property of $v$, and thus it can be ignored (recall that the theoretic distance between two suns is greater than two). If an unassigned vertex $u$ receives one or more offer messages originated by the same sun $s$, then $u$ turns its state to moon and stores the ID of $s$ in its system-sun property. Furthermore, $u$ stores the ID of all planets that forwarded the above offer messages in a property called \emph{system-planets}. This property is used by each moon $u$ to keep track of the planets adjacent to $u$ and in the same solar system as $u$. Finally, $u$ sends a confirmation message to its sun $s$ through a two-hop message (that requires two further supersteps to be delivered), which will be sent to one of the planets stored in the system-planets property. If $u$ receives offer messages from distinct vertices, then the above procedure is applied only for those messages originated by the sun $s$ with greatest ID. For every offer message originated by a sun $t$ with ID lower than the one of $s$, $u$ informs both $s$ and $t$ of the conflict through ad-hoc two-hop messages. These messages will be used by $s$ and $t$ to maintain a suitable data structure containing the information of each path between $s$ and $t$. At the end of this phase, all the galaxies of the generated sun vertices have been created and have diameter at most four. Also, some of the inter-system links have already been discovered, and this information will be useful in the following. The two steps described above are repeated until there are no more unassigned vertices. An example is illustrated in Fig.~\ref{fi:co}.

\smallskip\noindent{\bf Inter-System Link Generation.} In the first superstep, every planet and every moon broadcasts an \emph{inter-link discovery message} containing  the ID stored in the system-sun property of the vertex. In the next superstep, each vertex $v$ processes the received messages. All messages originated by vertices in the same solar system are ignored. Similarly as in the previous step, for each inter-link discovery message originated from a sun $t$ different from the sun $s$ of $v$, vertex $v$ informs both $s$ and $t$ of the conflict through two-hop messages that will be used by  $s$ and $t$ to maintain a suitable data structure containing the information of each path between $s$ and $t$. Once all messages have been delivered, each sun $s$ is aware of all links between its solar system and other systems. Also, for each link, $s$ knows what planet and moon (if any) are involved.  

\smallskip\noindent{\bf Next Level Generation.} 
In the first superstep, every sun $s$ creates a vertex $v_s$ whose level equals the level of $s$ plus one, and whose mass equals the sum of the masses of all the vertices in the solar system of $s$. Also, an \emph{inter-level edge} between $s$ and $v_s$ is created and will be used in the placement phase. In the next superstep, every sun $s$ adds an edge between $v_s$ and $v_t$, if $t$ is a sun of a solar system for which there are $k>0$ inter-system links. The edge $(v_s,v_t)$ is equipped with a \emph{weight} equal to the maximum number of vertices involved in any of the $k$ links. Finally, all vertices (except the newly created ones) deactivate themselves. 

\subsection{Placement Phase: \textsc{Distributed Solar Placer}}\label{sse:placing}

We now describe a \textsc{Distributed Solar Placer} algorithm, which behaves similarly to the \textsc{Solar Placer} of \fmmm.  After the coarsening phase, the only active vertices are those of the coarsest graph $G_k$. For this graph, the placement phase is not executed, and the computation goes directly  to the single-level layout phase (described in the next subsection). The output of the single-level layout phase is an assignment of coordinates to all vertices of $G_k$. Then, the placement phase starts and its execution is as follows.

In the first superstep, every vertex broadcasts its coordinates. In the second superstep, all vertices whose level is one less than the level of the currently active vertices activate themselves, and hence will start receiving messages from the next superstep. In the same superstep,  every vertex $v$ forwards the received messages to the corresponding vertex $v^*$ of lower level through its inter-level edge. Then $v$ deletes itself. At the next superstep, if a vertex $s$  receives a message, then $s$ is the sun of a solar system. Thanks to the received messages, $s$ becomes aware of the position of all suns of its neighboring solar systems. Hence, $s$ exploits this information (and the data structure containing information on the inter-system links), to compute the coordinates of all planets and moons in its solar system, as for the \textsc{Solar Placer}. Once this is done, $s$ sends to every planet $u$ of its solar system the coordinates of $u$. The coordinates of the moons are delivered through two-hop messages (that is, sent to planets and then forwarded). 

\subsection{Single-level Layout Phase: The \gila Algorithm}\label{sse:layout}

This phase is based on the \gila algorithm, the distributed single-level force-directed algorithm described in Section~\ref{se:preliminaries}. Recall that the execution of \gila is based on a set of parameters, whose tuning affects the trade-off between quality of the drawing and speed of the computation. The most important parameter is the maximum graph theoretic distance $k$ between pairs of vertices for which the pairwise repulsive forces are computed. Also, there are further parameters that affect the maximum displacement of a vertex, at a given iteration of the algorithm. The idea is to tune these parameters in order to achieve better quality for the coarser graphs, and shorter running times for the graphs whose size is closer to the original graph. Here we only describe how the parameter $k$ has been experimentally tuned, since it is the parameter that mostly affect the trade-off between quality and running time. The other parameters have been set similarly. For the drawing of every graph $G_i$, the value of $k$ is $6$ if the number of edges $m_i$ of $G_i$ is below $10^3$, it is $5$ if $10^3 \leq m_i < 5 \cdot 10^3$, it is $4$ if $5 \cdot 10^3 \leq m_i < 10^4$, it is $3$ if $10^4 \leq m_i < 10^5$, it is $2$ if $10^5 \leq m_i < 10^6$, and it is $1$ if $m_i \geq 10^6$.

\section{Experimental Analysis}\label{se:experiments}

We executed an experimental analysis whose objective is to evaluate the performance of \multi. We aim to investigate both the quality of the produced drawings and the running time of the algorithm, also in terms of scalability when we increase the number of machines. We expect that \multi computes drawings whose quality is comparable to that achieved by centralized multilevel force-directed algorithms. This is because the locality-based approximation scheme adopted by \gila (used in the single-level layout phase) should be mitigated by the use of a graph hierarchy. Also, we expect \multi to be able to handle graphs with several million edges in tens of minutes on an inexpensive PaaS infrastructure. Clearly, the use of a scalable vertex-centric distributed framework adds some unavoidable overhead, which may make \multi not suited for graphs whose size is limited to a few hundred thousand of edges. Our experimental analysis is based on three benchmarks called \regularb, \realb, and \bigb, described in the following.

%%%% REGULAR BENCHMARK %%%%%

\begin{table}[t]
\centering
\renewcommand{\arraystretch}{1.1}
  \caption{\small \regularb: number of vertices ($n$), number of edges ($m$), average number of crossings per edge (CRE), normalized edge length std deviation (NELD).}\label{ta:regularb}
\resizebox{\textwidth}{!}{\begin{tabular}{| l | r | r | c | c | c | c | l | r | r | c | c | c | c |}
    \hline
     \multicolumn{3}{|c|}{} & \multicolumn{2}{c|}{\fmmm} & \multicolumn{2}{c|}{\multi} & \multicolumn{3}{|c|}{} & \multicolumn{2}{c|}{\fmmm} & \multicolumn{2}{c|}{\multi}\\\hline
    \textsc{Name} & $n$ & $m$ & \textsc{CRE} & \textsc{NELD} & \textsc{CRE} & \textsc{NELD} & \textsc{Name} & $n$ & $m$ & \textsc{CRE} & \textsc{NELD} & \textsc{CRE} & \textsc{NELD} \\\hline
karateclub&34&78&1.10&0.25&1.09&0.33&Grid\_40\_40\_df&1,597&3,120&0.19&0.23&0.20&0.33
\\
snowflake\_A&98&97&0.00&0.25&0.11&0.21&Grid\_40\_40\_sf&1,599&3,120&0.39&0.18&0.38&0.31
\\
spider\_A&100&160&3.06&0.24&2.86&0.27&ug\_380&1,104&3,231&25.68&0.64&13.47&0.96
\\
cylinder\_010&97&178&0.35&0.16&0.72&0.08&esslingen&2,075&5,530&19.89&0.41&34.18&0.53
\\
sierpinski\_04&123&243&0.00&0.25&0.00&0.22&uk&4,824&6,837&0.07&0.36&0.06&0.65
\\
tree\_06\_03&259&258&0.40&0.29&1.54&0.17&4970&4,970&7,400&0.01&0.23&0.01&0.46
\\
rna&363&468&0.04&0.24&0.06&0.50&add20&2,395&7,462&60.38&0.50&100.44&0.50
\\
protein\_part&417&511&1.20&0.33&1.73&0.50&dg\_1087&7,602&7,601&0.06&0.34&0.00&1.04
\\
516&516&729&0.09&0.13&0.18&0.44&tree\_06\_05&9,331&9,330&8.63&0.47&19.65&0.93
\\
Grid\_20\_20&400&760&0.00&0.13&0.00&0.23&add32&4,960&9,462&1.31&0.88&0.97&1.66
\\
Grid\_20\_20\_df&397&760&0.24&0.23&0.20&0.34&snowflake\_C&9,701&9,700&0.00&0.64&0.00&0.40
\\
Grid\_20\_20\_sf&397&760&0.41&0.17&0.41&0.26&flower\_005&930&13,521&48.76&0.61&45.24&0.61
\\
dg\_617\_part&341&797&10.57&0.30&16.61&0.36&3elt&4,720&13,722&0.40&0.35&0.27&0.60
\\
snowflake\_B&971&970&0.00&0.42&0.00&0.39&data&2,851&15,093&2.15&0.39&2.52&0.64
\\
tree\_06\_04&1,555&1,554&8.53&0.35&7.04&0.19&grid400\_20&8,000&15,580&0.02&0.22&0.24&0.89
\\
spider\_B&1,000&1,600&7.03&0.24&8.26&0.73&spider\_C&10,000&16,000&171.31&0.32&262.09&0.93
\\
grid\_rnd\_032&985&1,834&0.00&0.15&0.00&0.30&grid\_rnd\_100&9,499&17,849&0.00&0.16&0.00&0.34
\\
cylinder\_032&985&1,866&0.46&0.19&0.44&0.39&sierpinski\_08&9,843&19,683&0.09&0.44&0.03&0.70
\\
cylinder\_100&985&1,866&4.60&0.18&4.48&0.45&crack&10,240&30,380&0.00&0.26&0.00&0.42
\\
sierpinski\_06&1,095&2,187&0.06&0.34&0.03&0.63&4elt&15,607&45,878&0.52&0.39&0.30&0.62
\\
flower\_001&210&3,057&47.37&0.67&45.97&0.47&cti&16,840&48,232&10.19&0.39&10.26&0.71
\\
Grid\_40\_40&1,600&3,120&0.00&0.15&0.00&0.32&&&&&&&		
\\
    \hline
  \end{tabular}}
\end{table}

\begin{figure}[t]
\subfigure[3elt]{\includegraphics[width=.2\columnwidth]{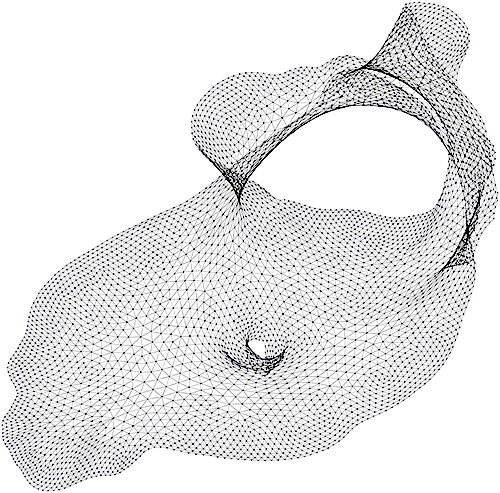}\hspace{3mm}\includegraphics[width=.22\columnwidth]{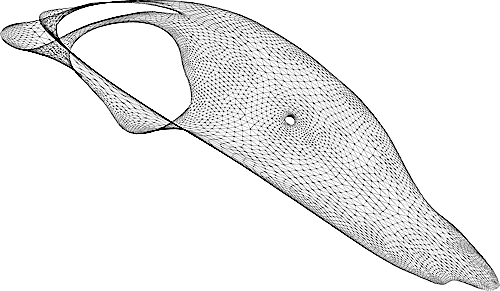}}\hfill
\subfigure[4elt]{\includegraphics[width=.2\columnwidth]{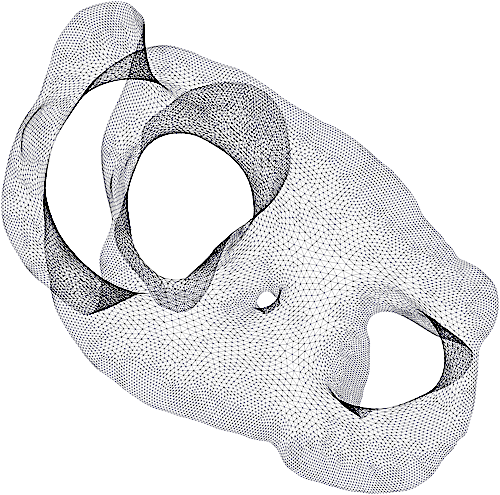}\hspace{3mm}\includegraphics[width=.22\columnwidth]{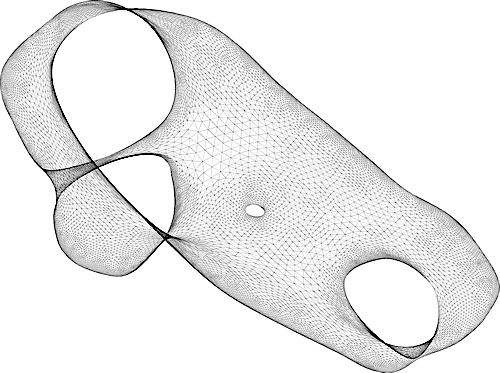}}
\subfigure[crack]{\includegraphics[width=.2\columnwidth]{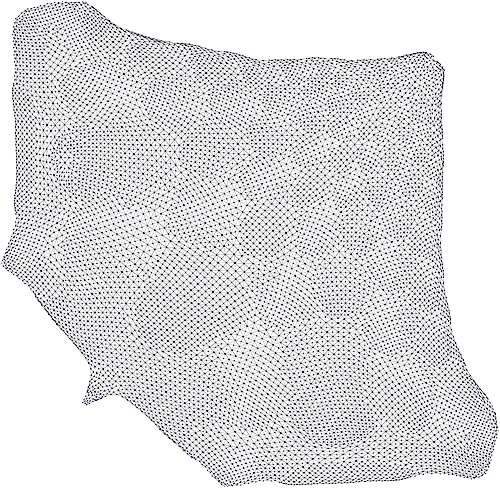}\hspace{3mm}\includegraphics[width=.22\columnwidth]{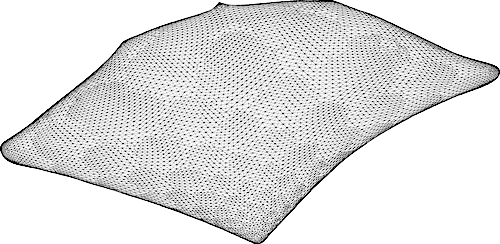}}\hfill
\subfigure[grid\_40\_40\_doublefolded]{\includegraphics[width=.2\columnwidth]{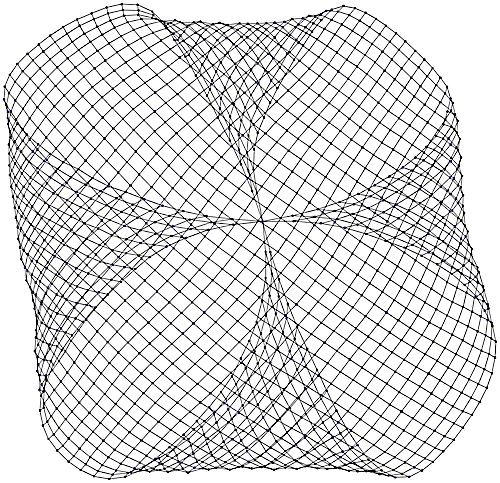}\hspace{3mm}\includegraphics[width=.16\columnwidth]{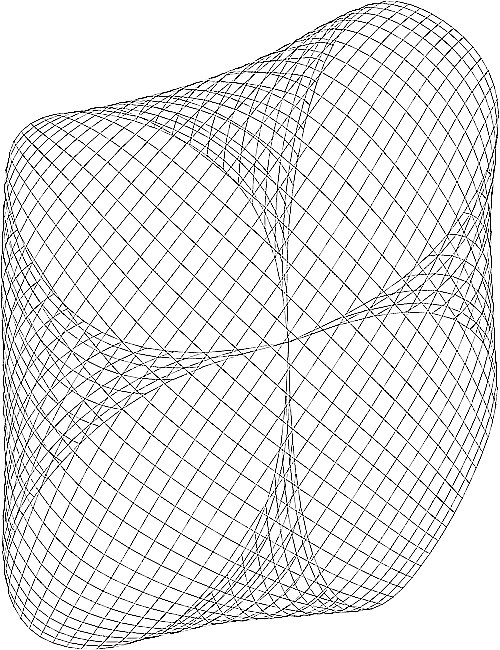}}
\subfigure[sierpinski\_06]{\includegraphics[width=.2\columnwidth]{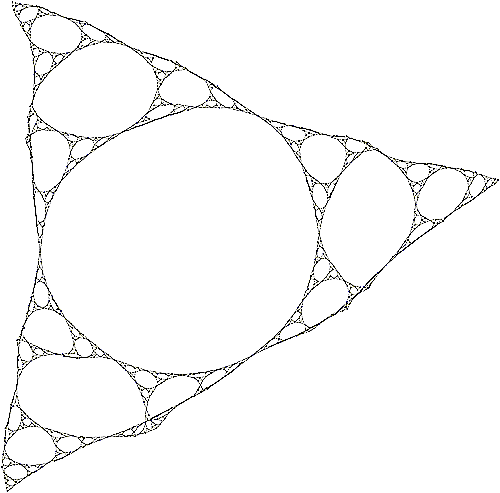}\hspace{3mm}\includegraphics[width=.12\columnwidth]{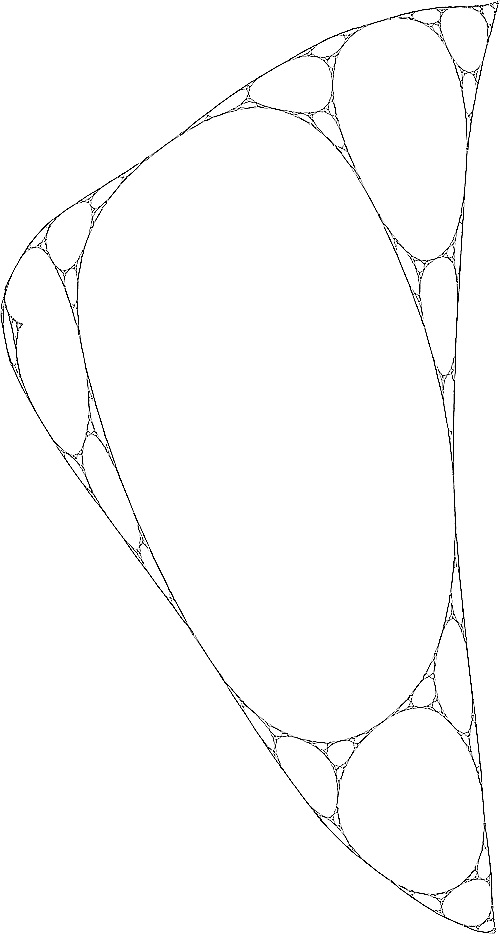}}\hfill
\subfigure[flower\_005]{\includegraphics[width=.2\columnwidth]{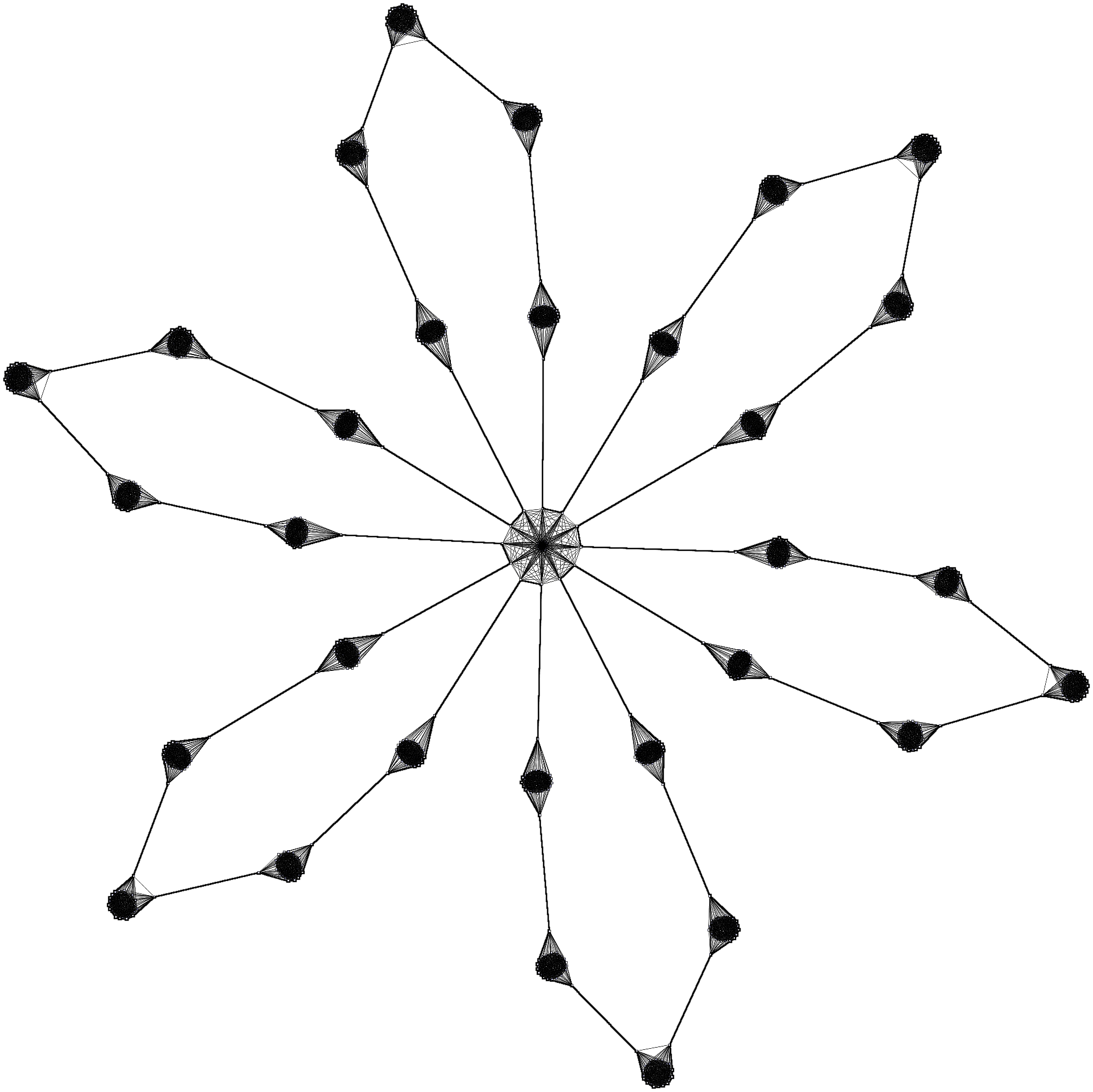}\hspace{3mm}\includegraphics[width=.2\columnwidth]{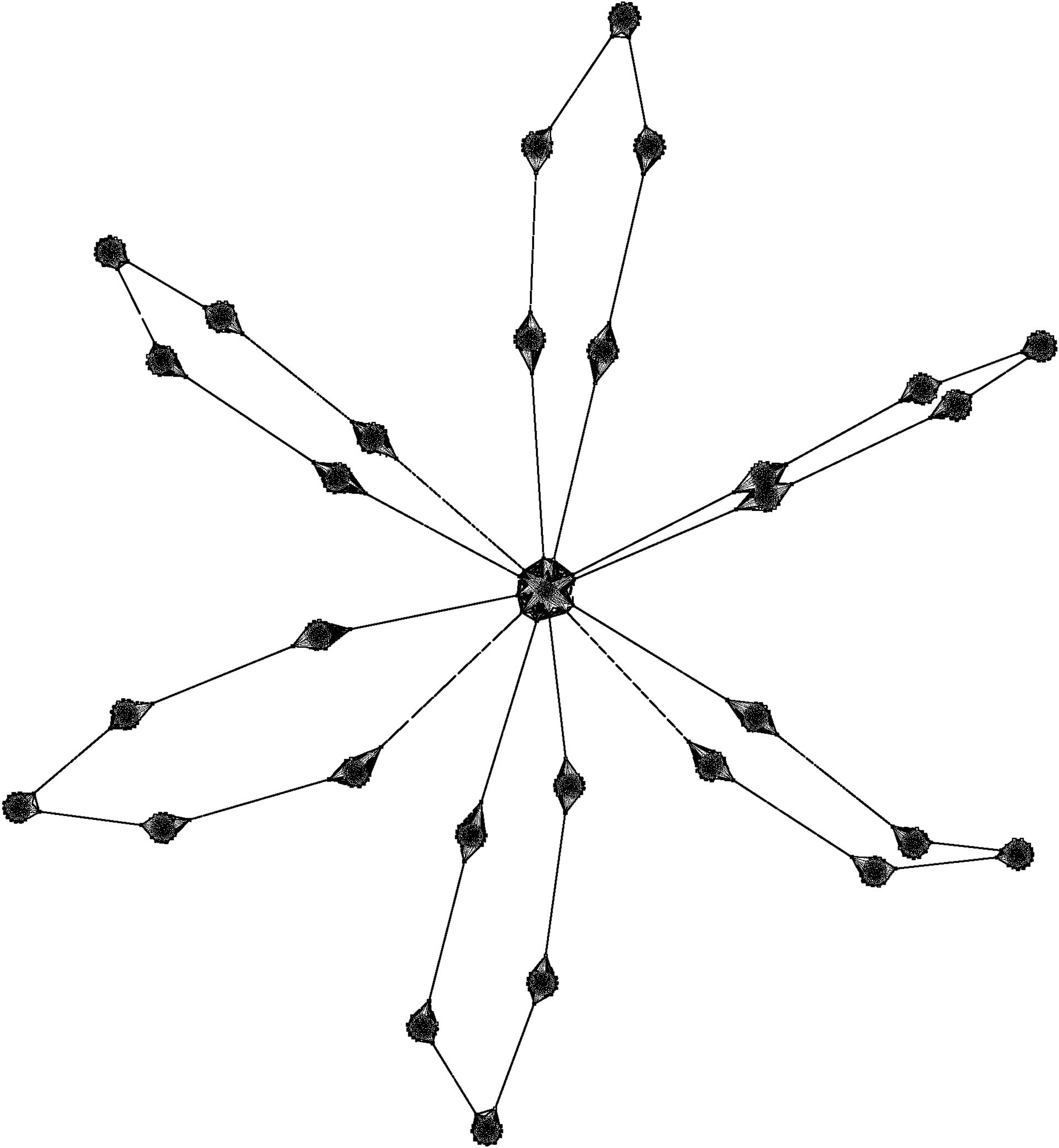}}
\caption{\label{fi:regular}Layouts of some \regularb instances. For each graph, the drawing computed by \fmmm (\multi) is on the left (right).}
\end{figure}

\begin{table}[t]
\centering
\renewcommand{\arraystretch}{1.1}
  \caption{\small Left: Details for  \realb. Right: Details for \bigb benchmark. Isolated vertices, self-loops, and parallel edges have been removed from the original graphs. The graphs are ordered by increasing number of edges.}\label{ta:details}
\resizebox{\textwidth}{!}{\begin{tabular}{| l | r | r | l | l | r | r | l |}
    \hline
    \textsc{Name} & $n$ & $m$  & \textsc{Description} & \textsc{Name} & $n$ & $m$ & \textsc{Description}\\\hline    
    asic-320 & 121,523 & 515,300 & circuit sim. problem & hugetric-10 & 6,600,000 & 10,000,000  & Mesh\\    
    amazon0302 & 262,111 & 899,792 & co-purchasing network & hugetric-20 & 7,100,000 & 10,700,000  & Mesh\\
    com-amazon & 334,863 & 925,872 & co-purchasing network & delaunay\_n22 & 4,100,000 & 12,200,000  & Triangulation\\
    com-DBLP & 317,080 & 1,049,866  & collaboration network &  &  &   & \\
    roadNet-PA & 1,087,562 & 1,541,514 & road network &  &  &   & \\
    \hline
  \end{tabular}}
\end{table}

\begin{table}[t]
\centering
\renewcommand{\arraystretch}{1.1}
  \caption{\small Running time of \multi on the \realb and \bigb instances, using increasing clusters of Amazon.}\label{ta:time}
\resizebox{\textwidth}{!}{\begin{tabular}{| l | c | c | c | l | c | c | c |}
    \hline
    &  \multicolumn{3}{c|}{Running time (seconds)}      &      &  \multicolumn{3}{c|}{Running time (seconds)} \\\hline
    \textsc{Name}  & 5 machines  & 10 machines  & 15 machines  & \textsc{Name} & 20 machines & 25 machines & 30 machines \\\hline    
    asic-320       &  1,626   &   1,102   &    1,281  & hugetric-10   & 7,923    & 4,828    & 3,679    \\
    amazon0302     &  2,518   &   2,696   &    1,577  & hugetric-20   & 9,891    & 8,243    & 4,445    \\
    com-amazon     &  3,400   &   3,395   &    2,242  & delaunay\_n22 & 8,160    & 3,301    & 3,932    \\
    com-DBLP       &  4,000   &   3,612   &    2,366  &               &          &          &          \\
    roadNet-PA     &  3,813   &   2,369   &    2,241  &               &          &          &          \\
    \hline
  \end{tabular}}
\end{table}

\begin{figure}[t]
\centering
\subfigure[]{\includegraphics[width=0.45\columnwidth]{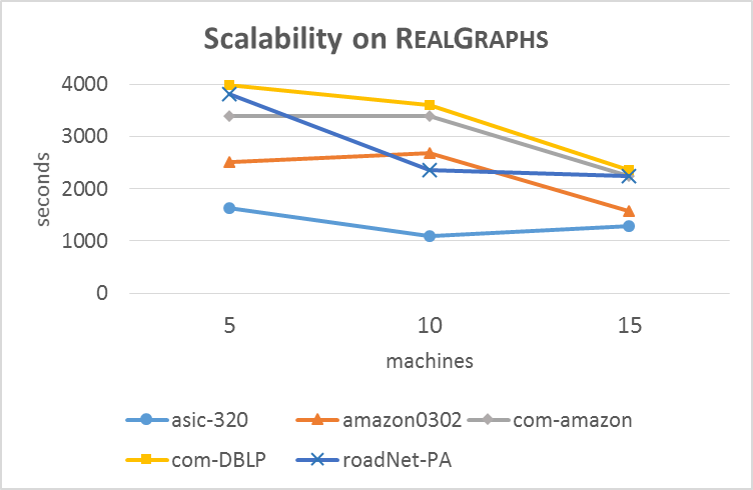}\label{ch:scal-real}}\hfil
\subfigure[]{\includegraphics[width=0.45\columnwidth]{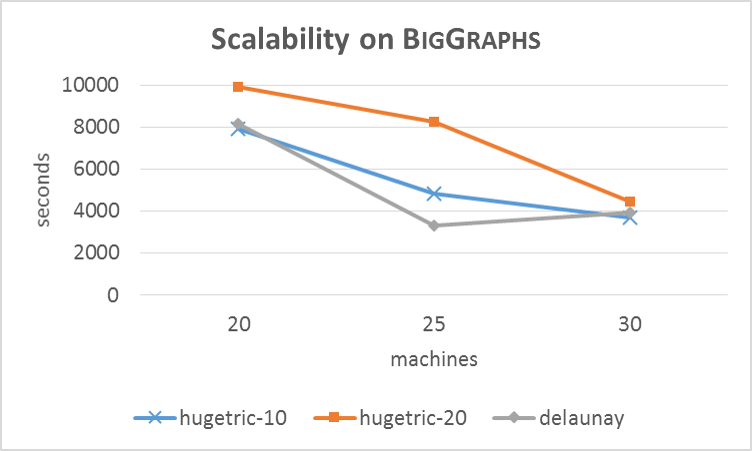}\label{ch:scal-big}}\hfil
\caption{Scalability of \multi on \realb instances.}\label{fi:scalability}
\end{figure}

The \regularb benchmark is the same used by Bartel et al.~\cite{DBLP:conf/gd/BartelGKM10} in an experimental evaluation of various implementations of the three main phases of a multilevel force-directed algorithm (coarsening, placement, and single-level layout). It contains $43$ graphs with a number of edges between $78$ and $48,232$, and it includes both real-world and generated instances~\cite{regular}. See also Table~\ref{ta:regularb} for more details. 
We used this benchmark to evaluate \multi in terms of quality of the computed drawings. Since the coarsening phase plays an important role in the computation of a good drawing, we first evaluated the performance of our \textsc{Distributed Solar Merger} in terms of number of produced levels compared to the number of levels produced by the \textsc{Solar Merger} of \fmmm. It may be worth remarking that, in the experimental evaluation conducted by Bartel et al.~\cite{DBLP:conf/gd/BartelGKM10}, the \textsc{Solar Merger} algorithm showed the best performance in terms of drawing quality when used for the coarsening phase. Our experiments show that the number of levels produced by the two algorithms is comparable and follows a similar trend throughout the series of graphs. The \textsc{Distributed Solar Merger} produces one or two levels less than the \textsc{Solar Merger} in most of the cases, and this is probably due to some slight difference in the tuning of the two algorithms. 
To capture the quality of the computed drawings, we compared \fmmm (the implementation available in the OGDF library~\cite{DBLP:reference/crc/ChimaniGJKKM13}) and \multi in terms of average number of crossings per edge (CRE), and normalized edge length standard deviation (NELD). The values of NELD are obtained by dividing the edge length standard deviation by the average edge length of each drawing. We chose \fmmm for this comparison for two main reasons: $(i)$ \multi is partially based on distributed implementations of the \textsc{Solar Merger} and of the \textsc{Solar Placer} algorithms; $(ii)$ \fmmm showed the best trade-off between running time and quality of the produced drawings in the experiments of Hachul and J{\"u}nger~\cite{JGAA-150}. The results of our experiments are reported in Table~\ref{ta:regularb}. The performance of \multi is very close to that of \fmmm in terms of CRE. In several cases \multi produces drawings with a smaller value of CRE than \fmmm (see, e.g., \texttt{ug\_380}). Concerning the NELD, \multi most of the times generates drawings with larger values than \fmmm. This may depend on how the length of the edges is set by the \textsc{Distributed Solar Placer} algorithm. However, also in this case the values of NELD follow a similar trend throughout the series of graphs. Fig.~\ref{fi:regular} shows a visual comparison for some of the graphs. Similarly to \fmmm, \multi is able to unfold graphs with a very regular structure and large diameter.

\begin{figure}[tb]
\centering
\subfigure[\texttt{asic-320}]{\includegraphics[width=0.16\columnwidth]{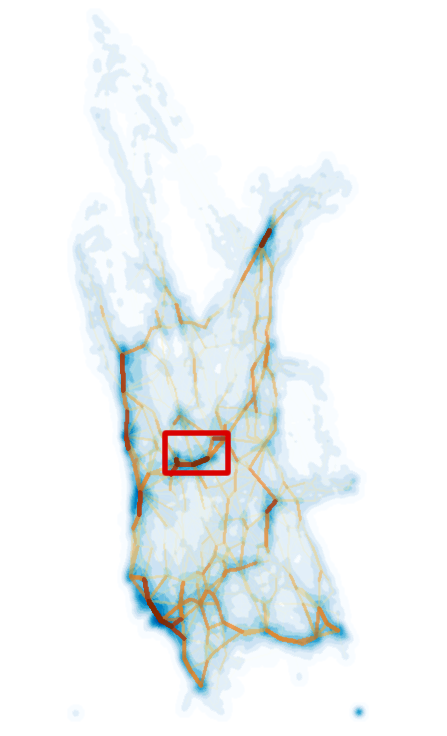}}\hfil
\subfigure[Detail of (a)]{\includegraphics[width=0.22\columnwidth]{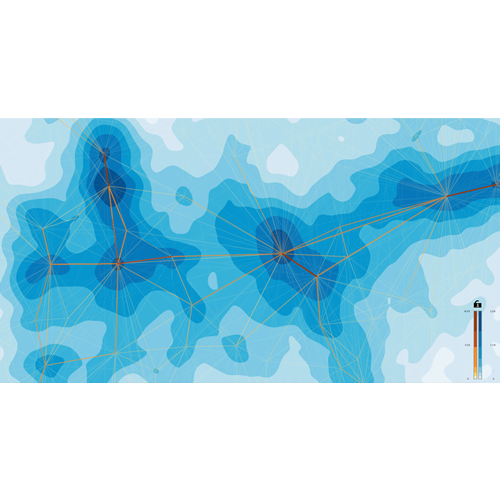}}\hfil
\subfigure[\texttt{com-amazon}]{\includegraphics[width=0.27\columnwidth]{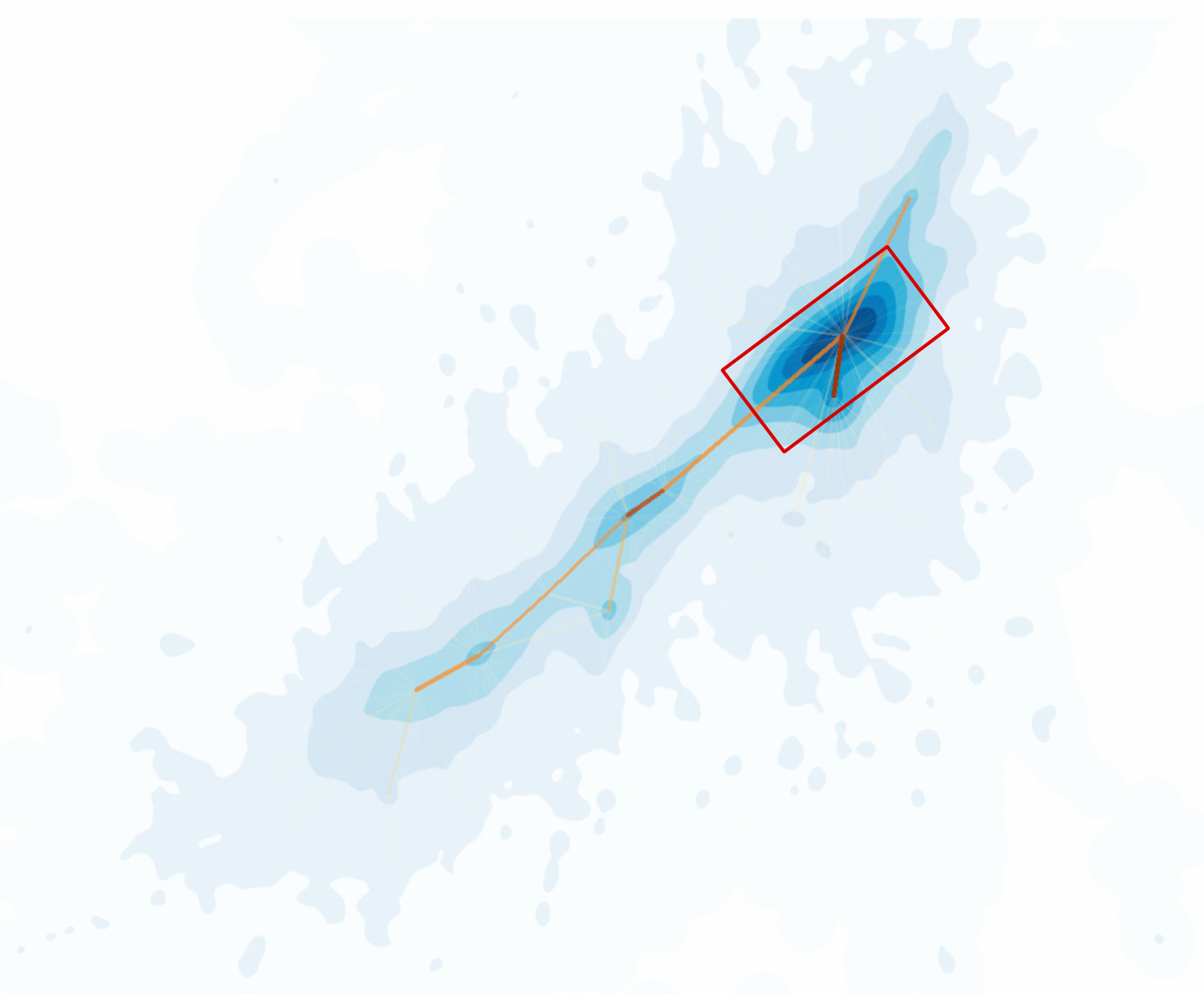}}\hfil
\subfigure[Detail of (c)]{\includegraphics[width=0.22\columnwidth]{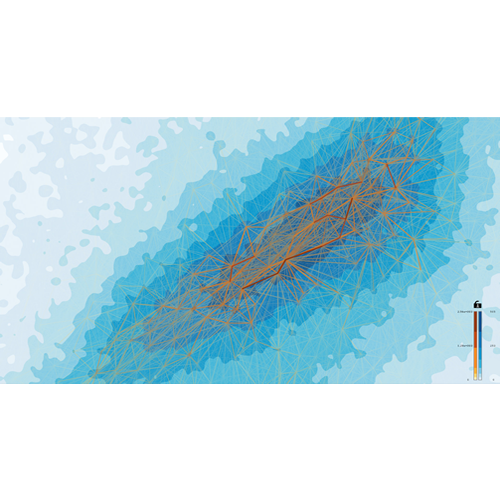}\vspace{5mm}}\hfil
\subfigure[\texttt{hugetric-10}]{\includegraphics[width=0.30\columnwidth]{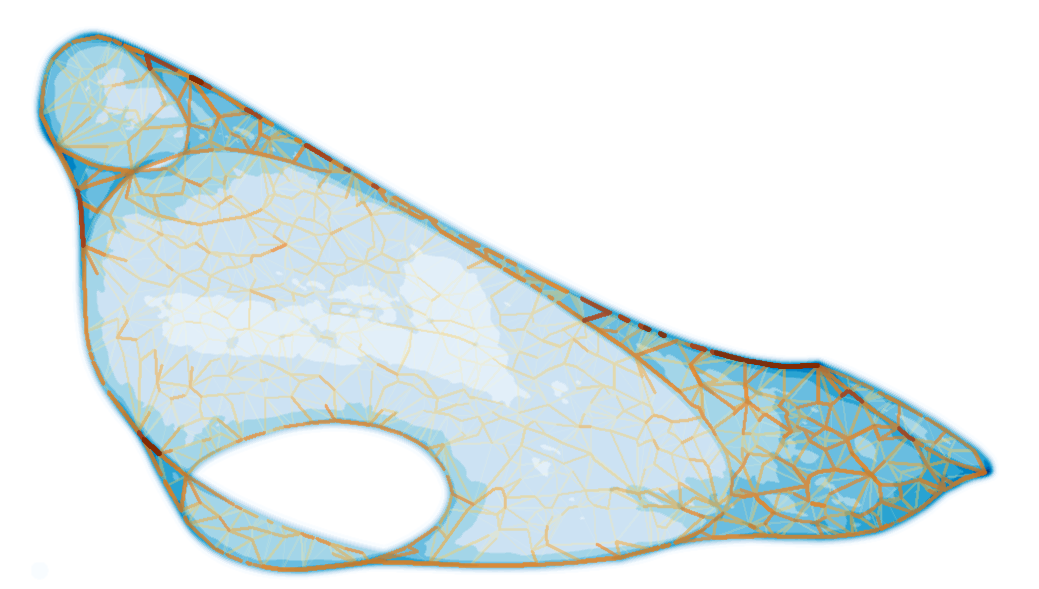}}\hfil
\subfigure[\texttt{hugetric-20}]{\includegraphics[width=0.30\columnwidth]{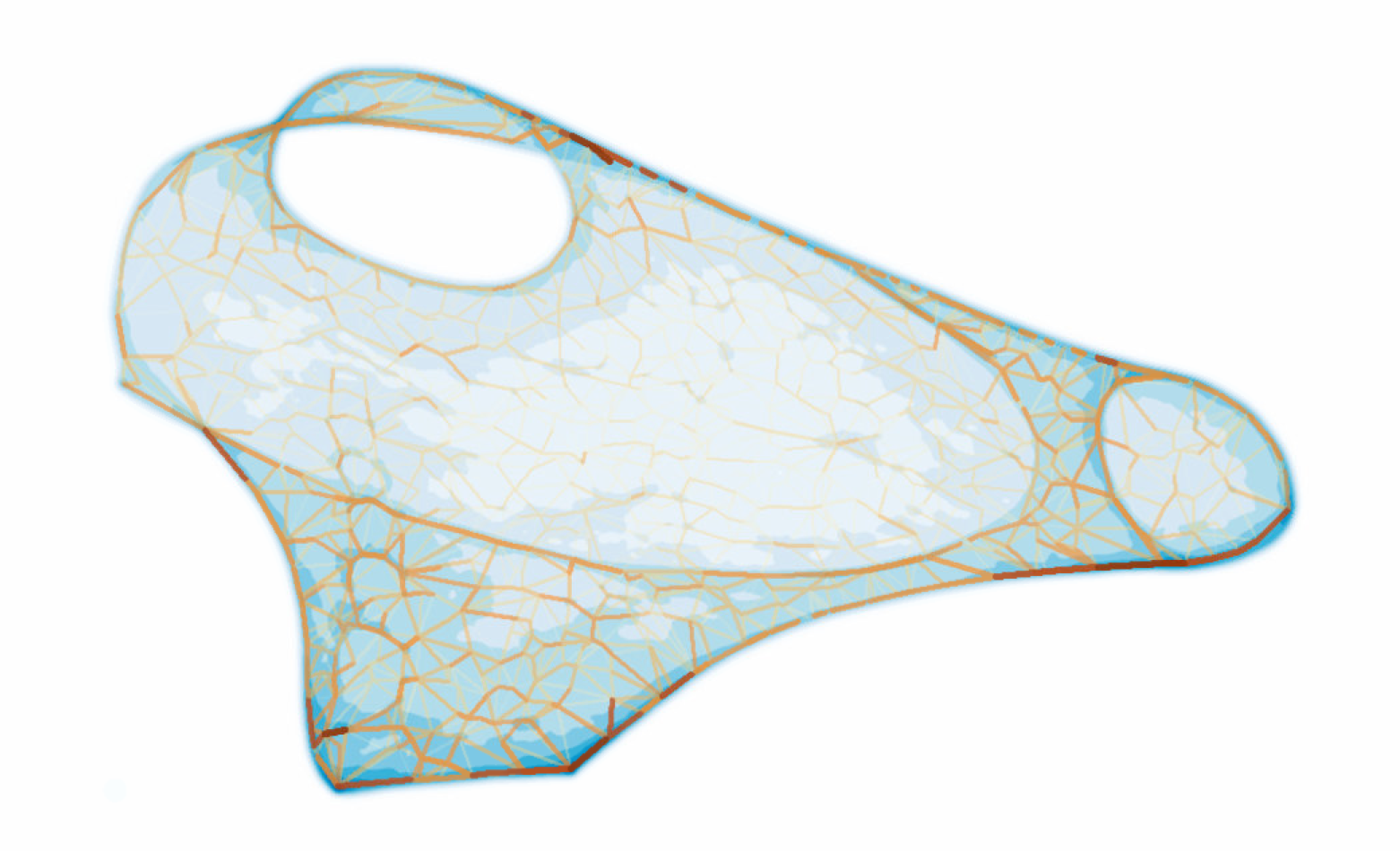}}\hfil
\subfigure[\texttt{delaunay\_n22}]{\includegraphics[width=0.22\columnwidth]{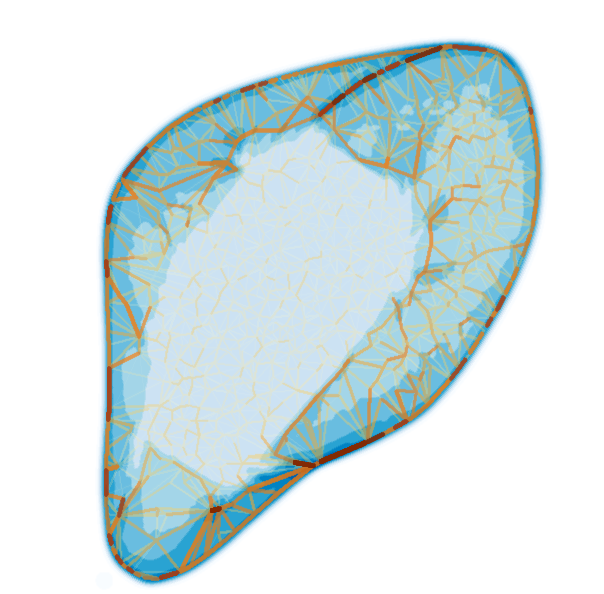}}
\caption{Layouts of (a--d) \realb instances and (e--f) \bigb instances computed by \multi and visualized (rendered) with \textsc{LaGo}.}\label{fi:ex}
\end{figure}

The \realb and \bigb sets contain much bigger graphs than \regularb, and are used to evaluate the running time of \multi, especially in terms of strong scalability (i.e., how the running time varies on a given instance when we increase the number of machines). The \realb set is composed of the $5$ largest real-world graphs (mainly scale-free graphs) used in the experimental study of \gila~\cite{arxiv}. These graphs are taken from the Stanford Large Networks Dataset Collection~\cite{snap} and from the Network Data Repository~\cite{network}, and their number of edges is between $121,523$ and $1,541,514$. The \bigb set consists of $3$ very large graphs with up to $12$ million edges, taken from the collection of graphs described in~\cite{nr-sigkdd16}\.%footnote{See also \url{http://www.networkrepository.com/}}.
Details about the \realb and \bigb sets are in Table~\ref{ta:details}. 

Table~\ref{ta:time} reports the running times of \multi on the \realb and \bigb instances, using increasing clusters of Amazon. Namely, for the \realb instances, $5$ machines were always sufficient to compute a drawing in a reasonable time, and using $15$ machines the time is reduced by $35\%$ on average. For the \bigb instances we used a number of machines from $20$ to $30$, and the reduction of the time going from the smallest to the largest cluster is even more evident than for the \realb set ($50\%$ on average). Fig.~\ref{fi:scalability} depicts the trend of the data in Table~\ref{ta:time}, showing the strong scalability of \multi. Fig.~\ref{fi:ex} shows some layouts of \realb and \bigb instances computed by \multi and visualized (rendered) with \textsc{LaGo}. 
It is worth observing that some centralized algorithm may be able to draw quicker than \multi graphs of similar size as those in the \realb set (see e.g.~\cite{DBLP:conf/gd/GodiyalHGH08}). This is partially justified by the use of a distributed framework such as Giraph, which introduces overheads in the computation that are significant for graphs of this size. However, this kind of overhead is amortized when scaling to larger graphs as those in the \bigb set. Also, using an optimized cluster rather than a PaaS infrastructure may improve the performance of the algorithm.

\section{Conclusions and Future Research}\label{se:future}

As far as we know, \multi is the first multilevel force-directed technique working in a distributed vertex-centric framework. Its communication protocol allows for an effective computation of a coarse graph hierarchy. Experiments indicate that the quality of the computed layouts compares with that of drawings computed by popular centralized multilevel algorithms and that it exhibits high scalability to very large graphs. Our source code is made available to promote research on the subject and to allow replicability of the experiments. In the near future we will investigate more coarsening techniques and single-level layout methods for a vertex-centric distributed environment.   

%Notes:
%\begin{itemize}
%\item Define the partitioning of the vertices for each coarse graph.
%\item Experiment further distributed single-level layout methods.
%\item Enlarge the experimental analysis with further multilevel algorithms (centralized, GPU-based,...).
%\item Experiment a filtering approach to make the input graph sparser.
%\end{itemize}

{\small \bibliography{multigila}}
\bibliographystyle{splncs03}

\clearpage

\appendix

\section*{Appendix}

\section*{Additional Material for Section~\ref{se:experiments}}

\begin{figure}
\centering
\includegraphics[width=0.9\columnwidth]{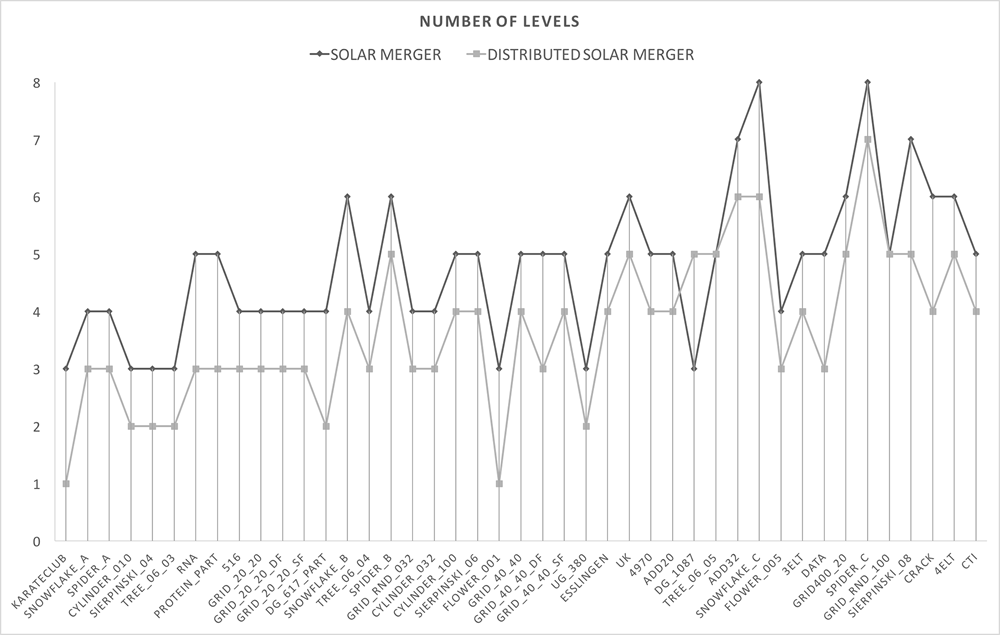}
\caption{\label{ch:levels}Number of levels produced by the \textsc{Solar Merger} and \textsc{Distributed Solar Merger} on the \regularb benchmark.}
\end{figure}

\begin{figure}
\centering
\subfigure{\includegraphics[width=0.9\columnwidth]{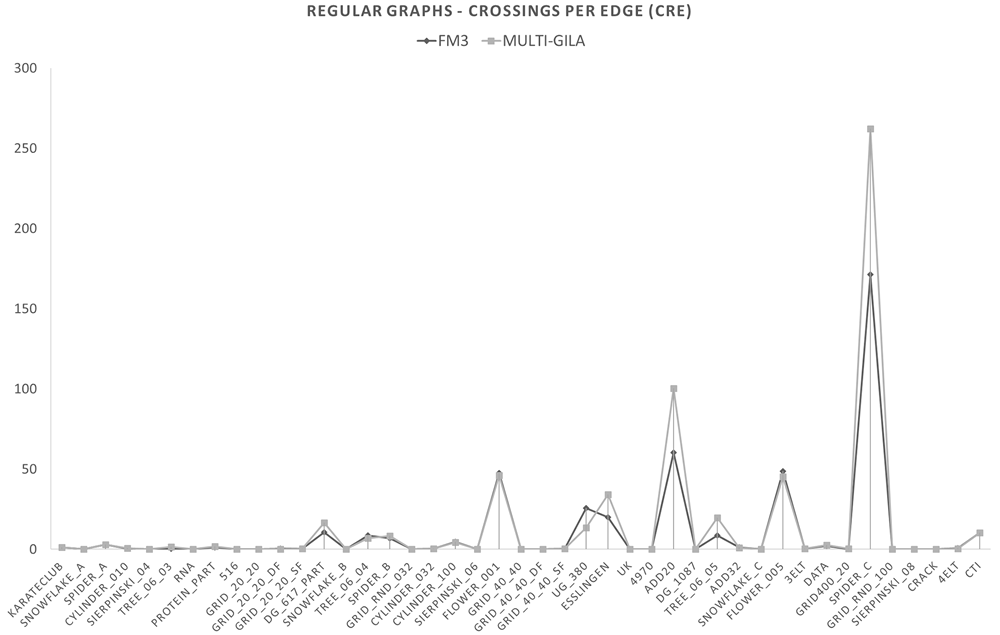}}
\subfigure{\includegraphics[width=0.9\columnwidth]{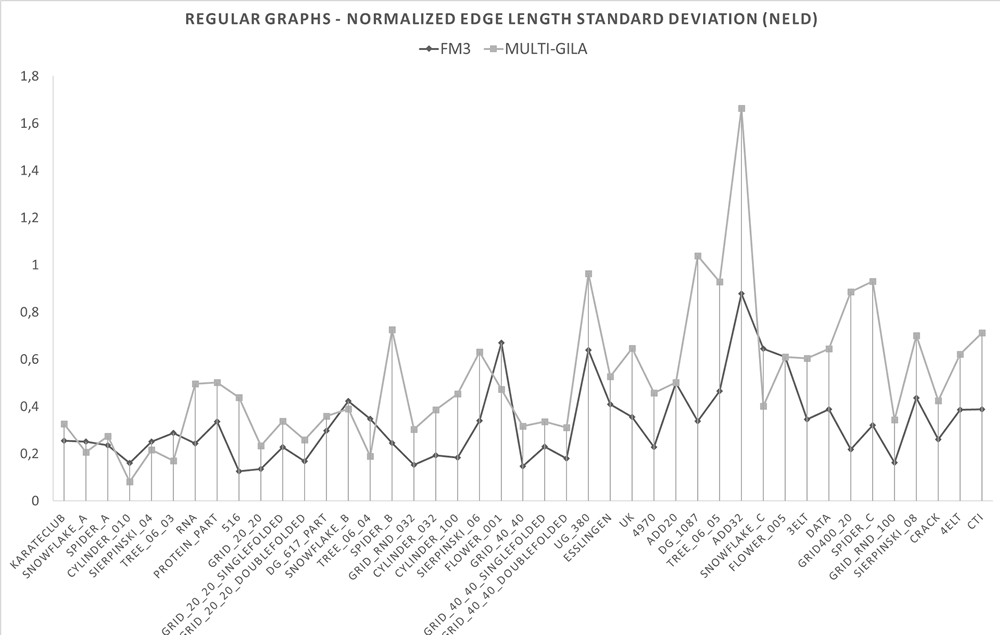}}
\caption{\label{ch:regular}Average number of crossings per edge (CRE) and normalized edge length standard deviation (NELD), for the \regularb benchmark.}
\end{figure}

\begin{figure}[t]
\centering
\subfigure[amazon0302]{\includegraphics[width=0.49\columnwidth]{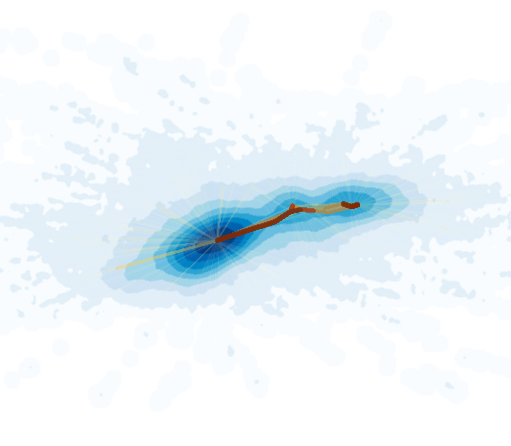}}\hfil
\subfigure[com-DBLP]{\includegraphics[width=0.49\columnwidth]{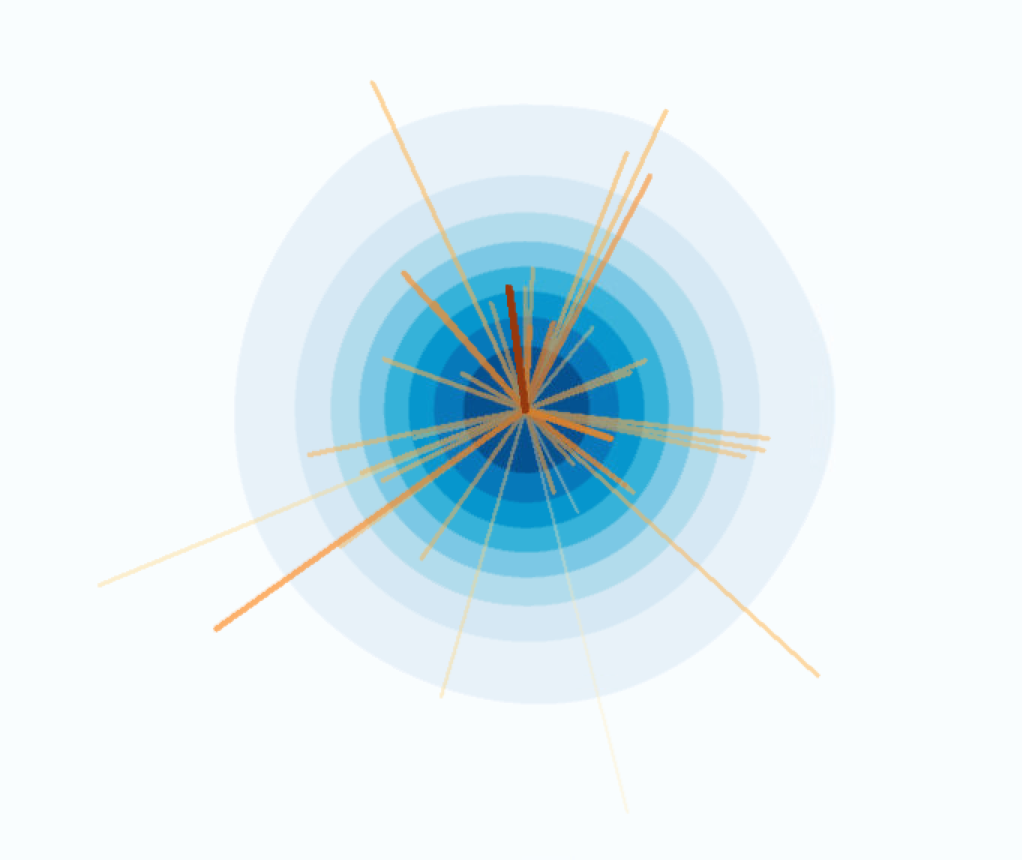}}
\subfigure[roadNet-PA]{\includegraphics[width=0.33\columnwidth]{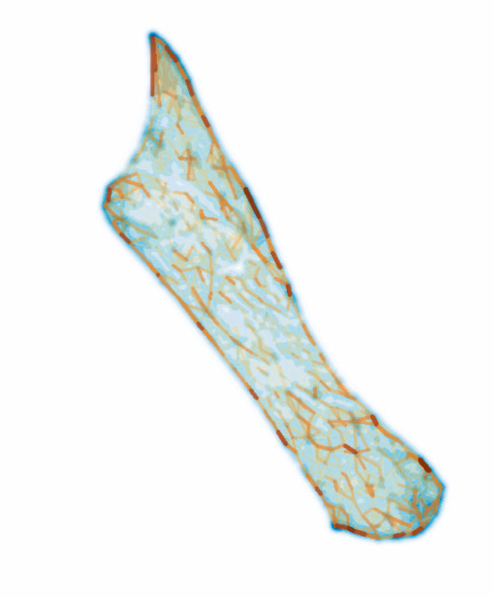}}
\caption{Additional layouts of \realb instances computed by \multi and visualized (rendered) with \textsc{LaGo}.}\label{fi:real-more-ex}
\end{figure}

\begin{figure}[t]
\centering
\subfigure[{\tt wiki-Talk}]{\includegraphics[width=0.45\columnwidth]{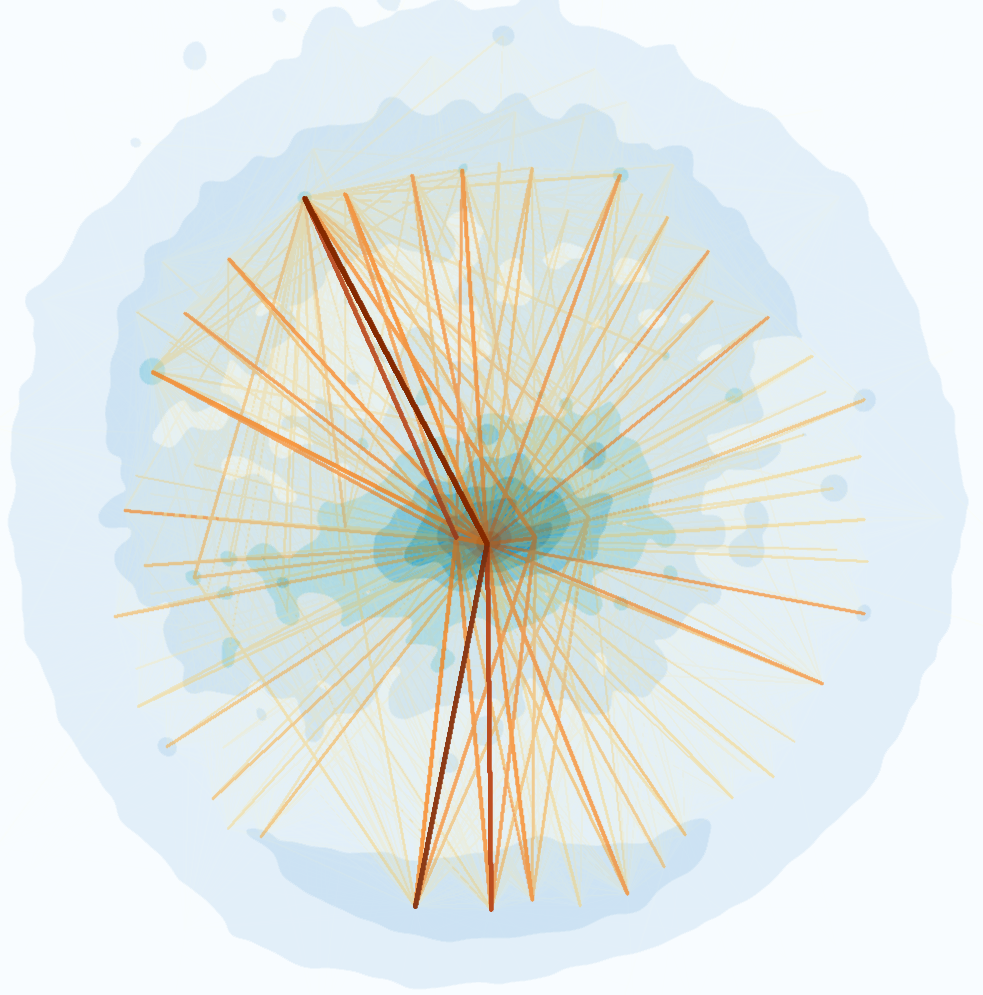}}\hfil
\subfigure[{\tt web-Google}]{\includegraphics[width=0.5\columnwidth]{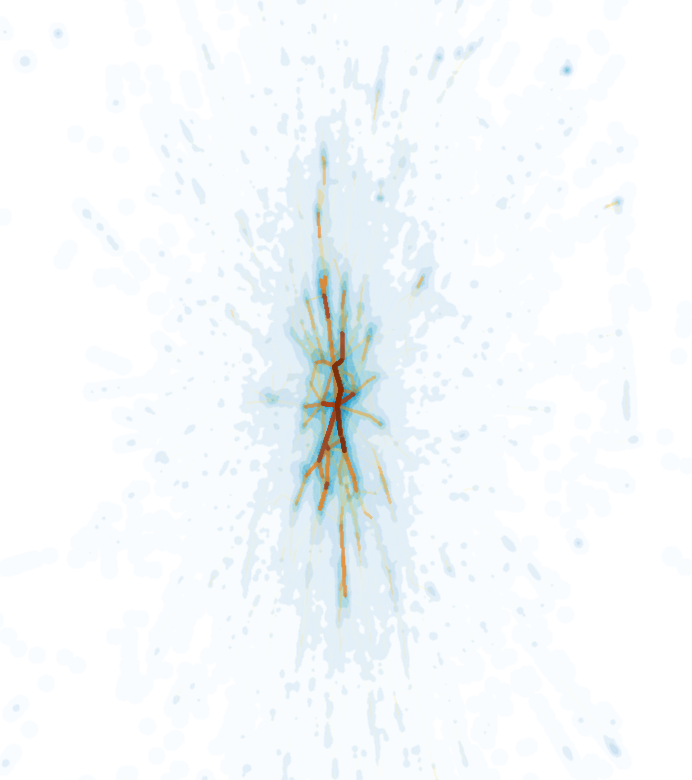}}
\caption{Additional layouts of big graphs computed by \multi and visualized (rendered) with \textsc{LaGo}: {\tt wiki-Talk} is a Wikipedia talk network with $2,390,000$ vertices and $5,020,000$ edges; {\tt web-Google} is a of a Google web graph with $880,000$ vertices and $5,100,000$ edges. On the Amazon cluster with $20$ machines, {\tt wiki-Talk} was drawn in $8,160$ seconds while {\tt web-Google} required $4,192$ seconds.}\label{fi:big-more-ex}
\end{figure}

\end{document}